\documentclass[twocolumn,prl,showpacs]{revtex4}%
\usepackage{graphicx}%
\usepackage{amsmath}%
\usepackage{amsfonts}%
\usepackage{amssymb}
\usepackage{bm}

\def\q{{ {\bf q} }}

\def\w{{\omega}}

\begin{document}
\title{
Reply to Comment on "Neutron-Inelastic-Scattering Peak 
by Dissipationless Mechanism in the $s_{++}$-wave State in Iron-based Superconductors" [arXiv:1106.2376] by
Y. Nagai and K. Kuroki}
\author{Seiichiro \textsc{Onari}$^{1}$,
and Hiroshi \textsc{Kontani}$^{2}$}
\date{\today }

\begin{abstract}

\end{abstract}

\address{
$^1$ Department of Applied Physics, Nagoya University and JST, TRIP, 
Furo-cho, Nagoya 464-8602, Japan. 
\\
$^2$ Department of Physics, Nagoya University and JST, TRIP, 
Furo-cho, Nagoya 464-8602, Japan. 
}
 

\sloppy

\maketitle
In 2009, we proposed the ``dissipationless mechanism'' \cite{onari-kontani}
to explain the hump structure in the neutron scattering spectrum 
Im$\chi^s(Q,\omega)$ in terms of the $s_{++}$-wave state:
While Im$\chi^s(Q,\omega)$ in the normal state is suppressed 
by the large inelastic scattering $\gamma(\w)$,
Im$\chi^s(Q,\omega)$ is increased to form a hump structure 
in the superconducting (SC) state since $\gamma(\w)=0$ for $|\w|<3\Delta$.
Later, Nagai and Kuroki revisited this issue
using {\it the same numerical method developed in Ref.} \cite{onari-kontani}, 
and claimed that the hump structure becomes small 
for a realistic gap size $\Delta=5$meV \cite{nagai-kuroki}.
However, their results fails to reproduce the 
particle-hole (p-h) gap $2\Delta$ in the spectrum,
which is a mathematical requirement at $T=0$.
After their report, we improved our numerical method that satisfies
this mathematical requirement for any $\Delta$, and
clarified that large hump appears even for $\Delta=5$meV \cite{onari-kontani2}.

This article is a reply to the comment by Nagai and Kuroki 
\cite{nagai-kuroki2} for our {unpublished paper}
in arXiv \cite{onari-kontani2}.
Their main claim is that 
(i) ``the main difference between Onari-Kontani's calculation
\cite{onari-kontani2} and ours \cite{nagai-kuroki} lies
in the choice of the quasiparticle damping in the normal
state $\gamma_0$, not in the method or the accuracy of the calculation.''
In Ref. \cite{onari-kontani2},
we put $\gamma_0=20$meV at $T=T_{\rm c}$
and $\gamma_s=10$meV in the SC state for $|\w|>4\Delta$ at $T=0$,
considering the thermal effect in the normal state. 
On the other hand, Nagai and Kuroki put $\gamma_0=\gamma_s=10$meV.
See details in Refs. \cite{onari-kontani2,nagai-kuroki2}.

\begin{figure}[h]
\includegraphics[width=0.8\linewidth]{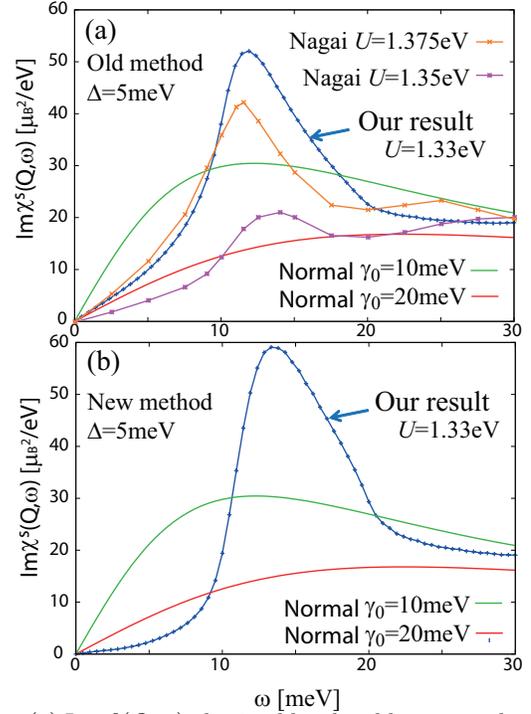}
\vspace{-5mm}
\caption{(a) Im$\chi^s(Q,\omega)$ obtained by the {old numerical method}
 \cite{onari-kontani} for $U=1.33$eV, in both the $s_{++}$-wave state 
with $\Delta=5$meV and $\gamma_s=10$meV and the normal state.
We also show the Nagai's results in Refs. \cite{nagai-kuroki,nagai-kuroki2}.
(b) Im$\chi^s(Q,\omega)$ obtained by the
{new numerical method} \cite{onari-kontani2}.
}
\label{fig2}
\vspace{-5mm}
\end{figure}


In Fig. \ref{fig2}, we show our numerical results 
given by the (a) old numerical method proposed in Ref. \cite{onari-kontani}
that was also used by Nagai and Kuroki \cite{nagai-kuroki,nagai-kuroki2},
and by the (b) improved method developed in Ref. \cite{onari-kontani2},
for $\Delta=5$meV, $\gamma_s=10$meV, $\gamma_0=10$ or 20meV, 
the cutoff energy $\Delta E=0.02$eV, and $U=1.33$eV.
The used model parameters are the same as those used by Nagai and Kuroki
 \cite{nagai-kuroki,nagai-kuroki2} except for $U$,
since we get the SDW state for $U=1.375$eV. 
{\it Contrary to the comment (i), 
we obtain large and clear hump structure 
in both (a) and (b) even when} $\gamma_s=\gamma_0=10$.
(In reality, $\gamma_0$ is always larger than $\gamma_s$ 
due to the thermal effect.)
The hump structure becomes prominent when the system 
is closer to the SDW state as shown in Ref. \cite{onari-kontani2}.
For $U=1.33$eV,
{\it the hump in the new numerical method (b) is apparently
larger than that in (a)}.
Comparing the peak positions in the normal states,
the present Stoner enhancement is smaller
than that in Nagai's result for $U=1.375$eV,



In Fig. \ref{fig2}, we also show Nagai's results 
by normalizing the scale to fix the value at $\w=0.03$eV:
We magnify Nagai's results for $U=1.375$ and $1.35$eV
by 0.42 and 0.57, respectively.
Based on these results, they claimed 
the smallness of the hump for $\Delta=5$meV.
However, even for $U=1.375$eV,
the hump structure is still smaller than ours 
with smaller Stoner enhancement,
while the p-h gap (=mathematical requirement at $T=0$) 
is shallower than ours.
Especially, in Fig. 1 (a) of Nagai's paper \cite{nagai-kuroki},
Im$\chi^s(Q,\omega)$ in the $s_{\pm}$-wave state for $|\w|<2\Delta$
is always larger than that in the normal state,
meaning that the p-h gap is failed to be reproduced.
Since the "dissipationless mechanism" is expected to 
work when $\gamma/\Delta\sim O(1)$
independently of the value of $\Delta$ \cite{onari-kontani2},
we consider that the Onari's results with larger hump are closer to the 
true numerical results, and
{Nagai's results might be insufficient for a quantitative
discussion, such as the size of the hump structure.}




In summary, contrary to Nagai-Kuroki's claim \cite{nagai-kuroki2}, 
we obtain a large hump structure in the $s_{++}$-wave state 
due to the ``dissipationless mechanism'',
even for $\gamma_0=\gamma_s\sim\Delta$.
As discussed in Ref. \cite{onari-kontani2}, 
we cannot distinguish between $s_{++}$- and $s_\pm$-wave states
by the spectrum at $\q=(\pi,\pi)$.

We are grateful to Kuroki and Nagai for the discussion
in the workshop ``Fe-based superconducutor'' 
at Yukawa Institute in Kyoto, on June 16-17, 2011.



\end{document}